# Photometric Study of Two Recently Discovered Variable Stars in the Field of BS Cas

Natalia A. Virnina[1], Radek Kocián[2], Ľubomír Hambálek[3], Pavol Dubovsky[4],
Ivan L. Andronov[1], Igor Kudzej[4]

1) Department "High and Applied Mathematics", Odessa National Maritime University, Ukraine, virnina@gmail.com, ilandronov@gmail.com
2) Observatory and Planetarium of Johann Palisa, VŠB - Technical University of Ostrava, 17. listopadu 15, 708 33 Ostrava-Poruba, Czech Republic, koca@astronomie.cz
3) Astronomical Institute, Slovak Academy of Sciences, 059 60 Tatranska Lomnica 18, Slovak Republic, hambalek@ta3.sk
4) Vihorlat Observatory, Mierova 4, Humenne, Slovak Republic, var@kozmos.sk,

**Abstract:** Two recently discovered variable stars (CzeV134 = GSC 3682 0018 = USNO-A2.0 1425-1870026 and CzeV135 = GSC 3682 2051 = USNO-A2.0 1425-1825909 = V1094 Cas), which have been identified in the field of the W UMa variable star BS Cas, are studied in the present paper. The phase curves and finding charts for these stars are presented. The ephemeris and other photometric parameters were computed.

The phenomenological features indicate that the first star (CzeV134) is probably a low-amplitude RRc Lyrae - type variable star with the period P = 0.419794±0.000029 d and the initial epoch $T_0$ = HJD2453236.50412±0.00056. The amplitude and the shape of the light curve are variable possibly indicating the Blazhko phenomenon. The second star (CzeV135) was classified as an EW-type binary system of subtype A. However, a β Lyrae type may not be excluded, as various classification parameters lie in a range of overlapping values for both classes. The period P=0.51429090±0.00000012 d and the initial epoch $T_0$ = HJD2454543.7920±0.0006. The O'Connell effect is clearly visible. There are slight changes of this effect, noticeable while comparing different seasons of observations. *O-C* diagrams for these stars were analyzed.

## Introduction

During the observation night 2007-08-22, when a time-resolving photometry of the variable star BS Cas was operated on the 200/1200 Newton telescope, two new variable stars were discovered (Hambálek 2008), using the function "Find variables" implemented in the program package C-Munipack (Motl, 2007).

The first new variable star CzeV134 = GSC 3682 0018 = USNO-A2.0 1425-1870026 at the position $\alpha(2000) = 01^h 22^m 26^s$, $\delta(2000) = +59° 12' 36''$ is already listed in the AAVSO database VSX of variable stars, however, the indicated type is EW (W UMa type). We suggest that it is a pulsating star, probably of the RRc type. However, the shape of the light curve is uncharacteristically instable: the amplitude of variability and the asymmetry of the phase curve are variable.

The second variable star CzeV135 = GSC 3682 2051 = USNO-A2.0 1425-1825909 is a close binary system at the position $\alpha(2000) = 01^h 20^m 23.373^s$, $\delta(2000) = +59°17' 15.05''$. The star is already listed in the GCVS (Samus' et al. 2011), named V1094 Cas and classified as an EW-type eclipsing variable. Based on a difference between the depths of minima, we suggest that it may be of the EB (β Lyrae) type.

In this paper, we make an analysis of all available light curves and correct classifications, ephemerids and other characteristics.

## Observations and photometric processing

The observations of the field of BS Cas, collected from three telescopes, were used for studying new variable stars:

1. 508/2500 Newton telescope of Stara Lesna Observatory (SL) equipped with SBIG ST-10XME CCD camera and *V*, *Rc*, *Ic* band filters, field of view (FOV) is 20′×13.5′, 1.82"/pixel (observer Hambálek L.)
2. 200/1200 Newton telescope of Johann Palisa Observatory and Planetarium in Ostrava (JP) with SBIG ST-8XME CCD camera and *V*, *Rc* and *Ic* band filters, FOV is 38.5′×25.5′, 3.02"/pixel (observer Kocián R.)
3. 280/1500 Newton telescope of observatory on Kolonica Saddle (KS), combined with Starlight Express SXVF-H9 camera and *Rc* filter, FOV is 20.5′×15.3′, 0.89"/pixel (observers Dubovsky P., Virnina N.)

Totally our observations cover a 6-year time interval from 2004-08-12 to 2010-08-09, which includes archive data, obtained before the new variable stars were discovered. The observations from the first telescope (SL) cover the interval from JD 2453230 to JD 2453990 (2004 and 2006), from the second telescope (JP) - JD 2454335 – 2455060 (2007 - 2009) with one additional night at JD 2455418 (2010), and from the third telescope (KS) - JD 2455052 – 2455061 (2009).

All the obtained CCD images were reduced by applying the bias, dark and flat field frames.

The light frames in *V, Rc* and *Ic* bands were obtained quasi-simultaneously (i.e. using alternately changing filters) using the SL and JP telescopes. However, the multi-color observations by the JP telescope were collected only in a couple of nights, and in the other nights only *Rc* filter was used.

For the photometry of both variable stars, the same three comparison stars were used. Initially, during the photometric process, these comparison stars were combined into an "artificial comparison star" with the magnitude equals to the unweighted mean magnitude of the chosen real comparison stars. This approach has two advantages. It improves the SNR of the "comparison star" because of larger number of counts, and it better accounts for any variations in transparency over the field, as opposed to using only one star. It also makes any anomalous reading immediately apparent.

This approach differs from that of the "CCD ensemble photometry on an inhomogeneous set of exposures" (Honeycutt 1992), where the weight for a given comparison star is estimated from the noise of the star, the sky and the detector.

Andronov and Baklanov (2004) and Kim, Andronov & Jeon (2004) used an "artificial comparison star" with weights being iteratively computed based on noise not only due to the factors mentioned above, but also from a possible variability of one or few comparison stars, i.e. based on an observed scatter. In their notation, there is a "main" comparison star, which is used for calibration. The weighted mean values of brightness of other comparison stars are free parameters determined from the iterations. This means that other stars are used to improve a number of counts (and one even does not need to know their magnitudes), but only one ("main" star) is used for the calibration. Such an approach does not produce systematic shifts of the mean brightness due to using different samples of the comparison stars. In this paper, we use "unweighted" comparison star for the "Var-Comp" values. However, for the calibration of our data, we use only one "main" star.

There is no reliable information about magnitudes of any stars in the 30′ vicinity of BS Cas in any photometric catalogs, thus we used *R* magnitude as an average of *R*1 and *R*2 measurements given in the catalog USNO-B1.0. The USNO-B1.0 numbers, the coordinates, and *R*1, *R*2 and average *R* magnitudes of the comparison stars are listed in Table 1. For the *V*- and *Ic*-bands the instrumental magnitudes were measured and shown in the present paper. The positions of comparison stars, as well as the positions of both variables, are marked on Fig. 1. The mean differences between comparison stars in the instrumental systems are: C1-C2=-0.096$^m$ (I-band, SL), C2-C3=-0.939$^m$ (I-band, SL), C1-C2=-0.070$^m$ (R-band, SL), C2-C3=-1.356$^m$ (R-band, SL), C1-C2=-0.035$^m$ (V-band, SL), C2-C3=-1.756$^m$ (V-band, SL), C1-C2=-0.051$^m$ (R-band, JP), C2-C3=-1.429$^m$ (R-band, JP), C1-C2=-0.048 (R-band, KS), C2-C3=-1.531 (R-band, KS).

After applying the transformation formulae, described in the next paragraph, to the *Rc*-band instrumental photometric measurements using the "unweighted artificial comparison star", we used the function RUN in the program MCV (Andronov & Baklanov, 2004) and chose the second comparison star as the "main" one for the calibration, with the magnitude, indicated in Table 1 in the column *R*.

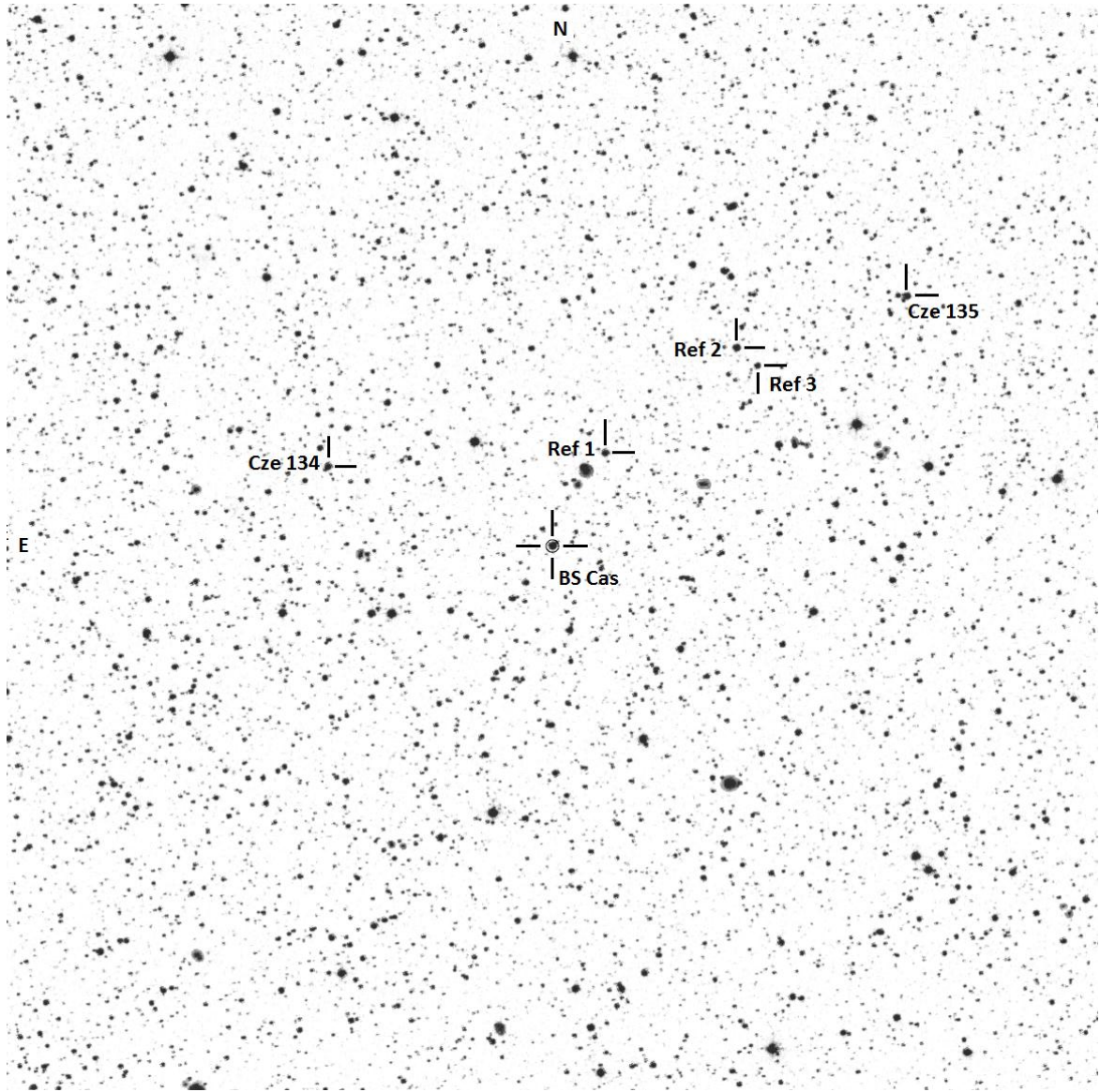

Fig. 1. 30′×30′ chart of the vicinity of BS Cas, both new variable stars and all three reference stars are marked. The "main" star used for the calibration, is Ref 2.

Table 1. Names, coordinates and *R* magnitudes of comparison stars

| # | USNO-B1.0 | coordinates | | R1 | R2 | R |
|---|-----------|-------------|---|------|------|------|
| 1 | 1492-0043777 | 01$^h$ 21$^m$ 27.376$^s$ | +59° 12′ 58.53″ | 11.860 | 11.830 | 11.845 |
| 2 | 1492-0043525 | 01$^h$ 20$^m$ 59.458$^s$ | +59° 15′ 51.65″ | 11.480 | 11.430 | 11.455 |
| 3 | 1492-0043467 | 01$^h$ 20$^m$ 55.011$^s$ | +59° 15′ 22.36″ | 12.800 | 12.990 | 12.895 |

## Photometric systems agreement

As different involved telescopes are equipped with different CCD cameras, and each chip has its own curve of quantum efficiency, and different photometric filters were used with these telescopes, the transformation coefficients should be found to match the observations on the phase curves. There are no standard magnitudes for any of the stars in the vicinity of BS Cas, as well as there are no photometric standards for nearby known variable stars, which could be used as the reference stars.

The photometry, obtained from the observations on the first (SL) telescope (508/2500 Newton) is the most accurate. Thus we chose these *V, Rc, Ic* bands as reference system and obtained the transformation formulae for other used photometric systems. The observations obtained on the SL telescope, cover the phase curves completely. The points collected from the JP telescope, nearly cover the phase curves only in *Rc*-band. As we are interested in the time-dependent changes in the shape of the phase curves, and only *Rc*-bands observations from each of three telescopes completely (or almost completely) cover the phase curves, the transformation coefficients for the *Rc* filters were found. For this we used 30 non-variable stars, including 3 comparison stars for the studied variables. The following transformation formulae were found for the instrumental magnitudes of the *Rc*-band measurements:

$$Rc^1 = 1.065(19) \cdot Rc^2 - 0.065(18) \cdot V^1 - 0.023(4),$$

and

$$Rc^1 = 1.095(16) \cdot Rc^3 - 0.095(14) \cdot V^1 - 0.007(3),$$

where the upper indices refer to the ordinal number of telescope. The linear trends' coefficients were found from the $(V^1 - Rc^2)$ vs $(V^1 - Rc^1)$ and $(V^1 - Rc^3)$ vs $(V^1 - Rc^1)$ diagrams using the linear polynomial approximation function of the MCV software (Andronov & Baklanov, 2004).

Here the zero-points correspond to an "unweighted artificial comparison star", whereas for the analysis we recompiled the data using the "weighted" method and C2 as a main comparison star.

## Photometric analysis and *O–C* diagrams

The preliminary basic elements for these two variable stars were found with the period analysis software Peranso v2.2 (Vanmunster, 2007). The Lafler-Kinman method (Lafler & Kinman, 1965) was used. All the parameters needed for the General Catalog of Variable Stars (GCVS, Samus et al. 2011) have been determined with corresponding error estimates for both stars.

All HJD photometry data are attached to the paper as appendices and are available from the OEJV web-site. The time used is UTC, taking into account the "leap second".

*CzeV134 = USNO-A2.0 1425-1870026* is a periodic pulsating variable star with the period P=0.419794(29) d, which was calculated using all observations in *Rc*-band from all involved telescopes, reduced to the reference photometric system as it was described above. As it is clearly visible in Fig. 2, different parts of the time-resolved light curve don't match on the combined phase curve. The shape of the light curve, the asymmetry and the amplitude of variability are variable, which is also visible in Fig. 2. This indicates a possible Blazhko phenomenon (see Chadid et al. 2010, Poretti et al. 2010 for recent studies). Our CCD images were reduced with the calibration frames and the photometry data was carefully corrected, thus the most probable reason of this effect is the presence of physical changes in the pulsations, which causes the variability of the phase curve. As the shape of maximum and asymmetry are variable, the first clearly observed maximum was chosen as the initial epoch. The maximum timing and its error estimate were calculated by smoothing the light curve by the algebraic polynomial of degree $n = 7$ (Andronov, 1994, 2003,

Mikulášek et al., 2006): HJD $T_{max}$=2453236.50412(56). On the phase curve, some maxima occur ahead of the phase zero, the others are delayed. The variability of maxima height is illustrated on Fig. 3 – the plot of time dependence of magnitudes in maxima. The maxima are asymmetric and the phase curves' shapes are variable, thus, to determine extrema timings, each observed maxima was fitted by the algebraic polynomial of degree $n = 6$ or $n = 7$ (Mikulášek et al., 2006). No reliable periodicity was found in this variability. All maxima timings with the time errors, $O$-$C$ and $Rc$ magnitudes are listed in Table 2. Using these data, the $O$-$C$ diagram was plotted in Fig. 4 with the corresponding error bars. These points could be fitted with the parabolic curve (MCV, Andronov & Baklanov, 2004):

$$O\text{-}C = 3.9(\pm 2.2)\cdot 10^9 \cdot E^2 - 2.4(\pm 1.0)\cdot 10^5 \cdot E + 0.0137(\pm 0.0098).$$

However, as neither of the coefficients is statistically significant, we cannot conclude that the changing of the period variation is real.

Using the FDCN program (Andronov, 1994, 2003) the trigonometric polynomial fit of the statistically optimal degree $s = 5$ was found. It yielded the average $Rc$-band magnitudes in minimum and maximum: $Rc_{min} = 11.633(2)$ mag, $Rc_{max} = 11.457(2)$ mag, respectively. The asymmetry of the smoothing trigonometric polynomial is 0.519(8), which could be interpreted as the average asymmetry of the real phase curve.

The magnitudes of the comparison stars in $V$ and $Ic$ – bands are unknown. But we've made instrumental photometric measurements in these bands. The phase curves are presented in Fig.5. Changes of the shape of the phase curves are visible in all 3 filters.

The instrumental indices $V$-$Rc$, $Rc$-$Ic$ and $V$-$Ic$ shown in Fig. 6 were calculated as well and their phase curves plotted with the same period and initial epoch as the phase light curve. The real magnitudes of the color indices are unknown because of absence of calibration of this field, but their changes are clearly visible. The color indices phase curves are sinusoidal, and they are stable, without any visible changes. Maximum of brightness of the star corresponds to the maximum on the color indices phase curves, i.e. the color temperature of the star is at maximum during the brightness maximum.

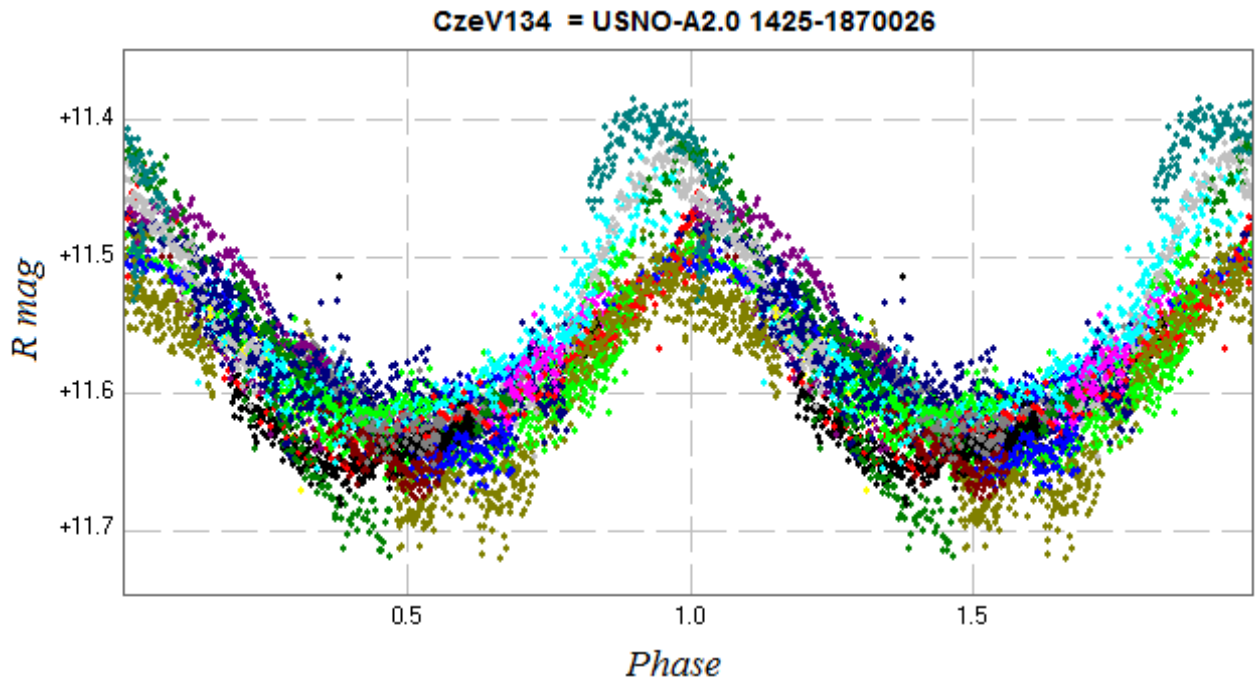

Fig. 2. The combined $Rc$ phase curve of pulsating variable star CzeV134. Different colors correspond to different nights.

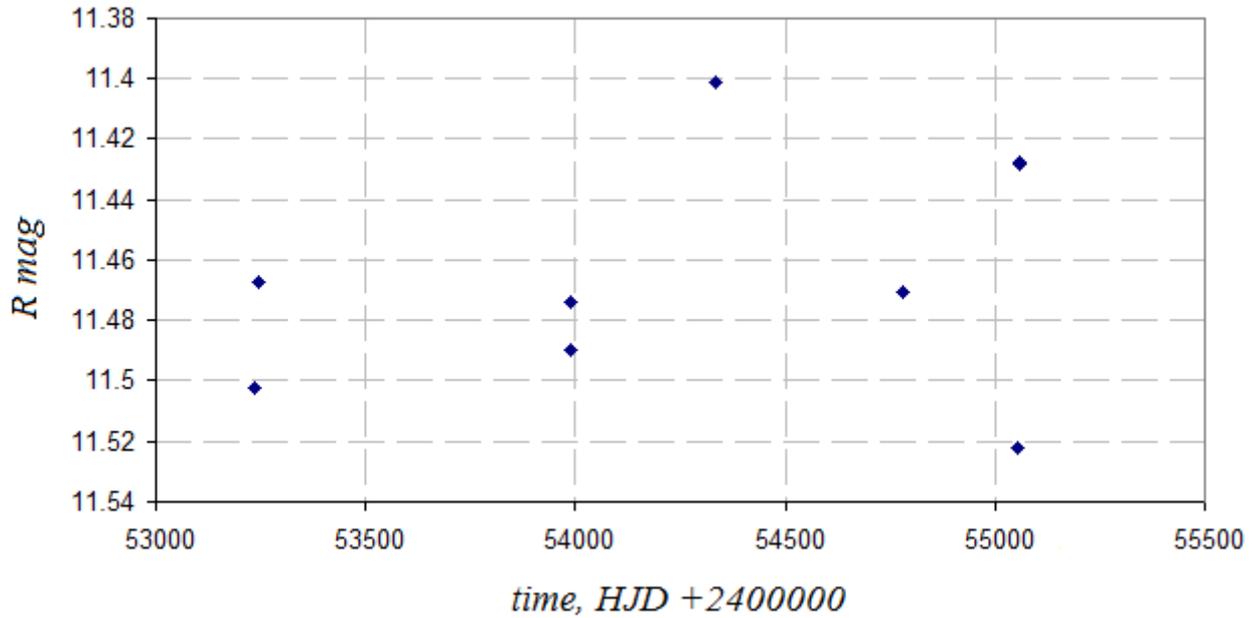

Fig. 3. Variations of the *Rc*-band magnitude of CzeV134 at maximum.

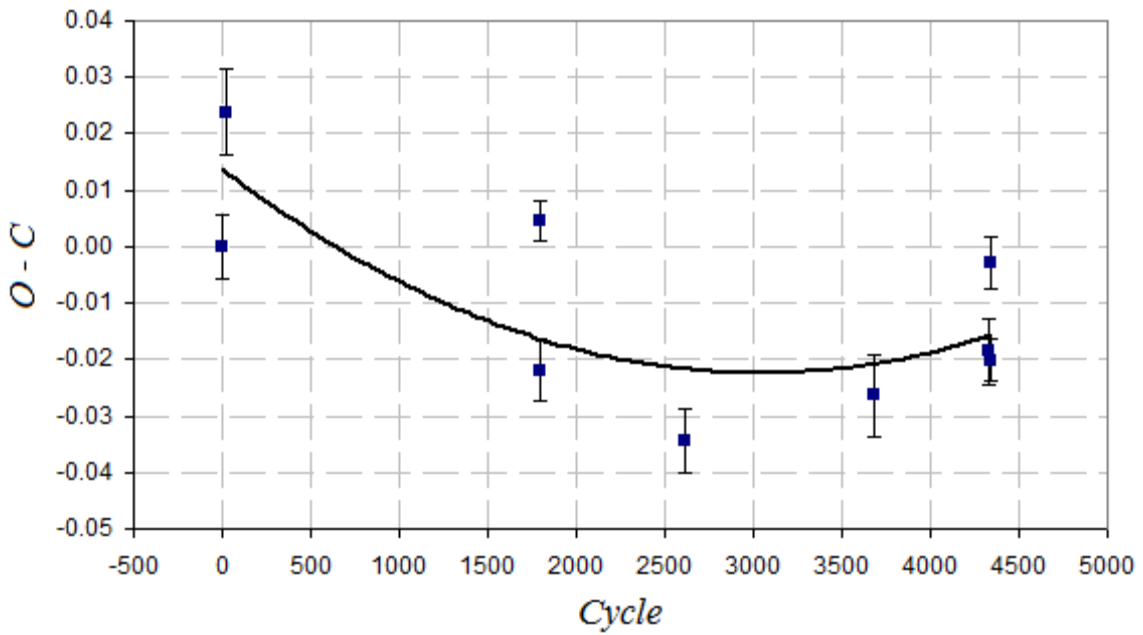

Fig. 4. *O-C* diagram for maxima for CzeV134.

*CzeV135 = USNO-A2.0 1425-1825909 = V1094 Cas* was found to be a periodic variable star as well. The preliminary period of variability $P = 0.514292(20)$ d was found using the Peranso software (Vanmunster 2007). According to the phenomenological classification given in GCVS, the type of variability is EB ($\beta$ Lyrae - type). After the approximate period was found using the method of Lafler & Kinman (1965), the program FDCN (Andronov, 1994, 2003) was used to compute the coefficients of the statistically optimal smoothing trigonometric polynomial of degree $s = 8$, using the least squares method routine and differential correction for the periods. The observations from all telescopes were used, and the following ephemerides were determined:

$$HJD\ T_{\min I} = 2454543.7936 + 0.51429150 \cdot E$$
$$\pm 0.0004 \quad \pm 0.00000011$$

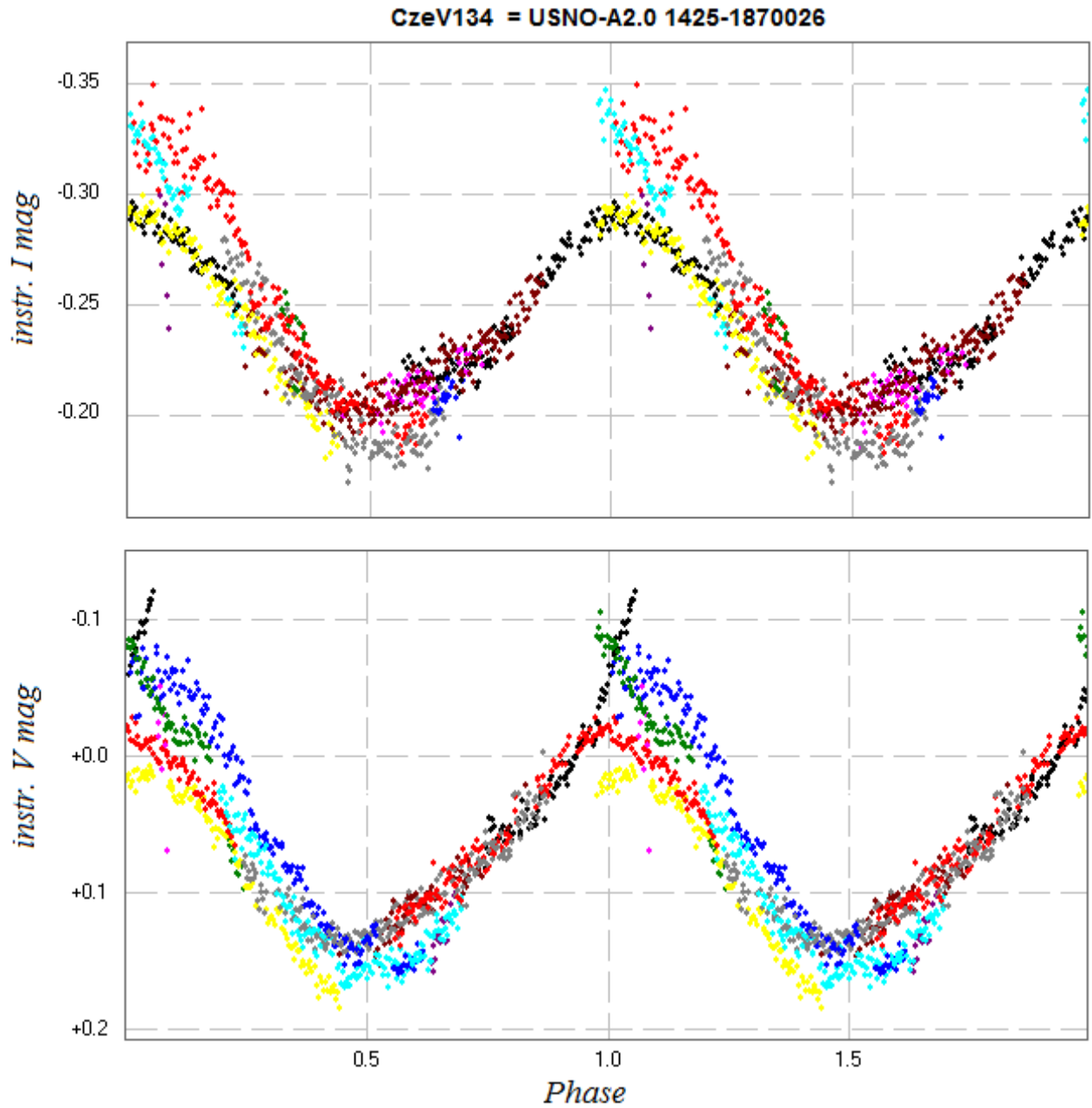

Fig. 5. Instrumental *V*- and *Ic*-band phase curves for CzeV134. Different colors correspond to different nights.

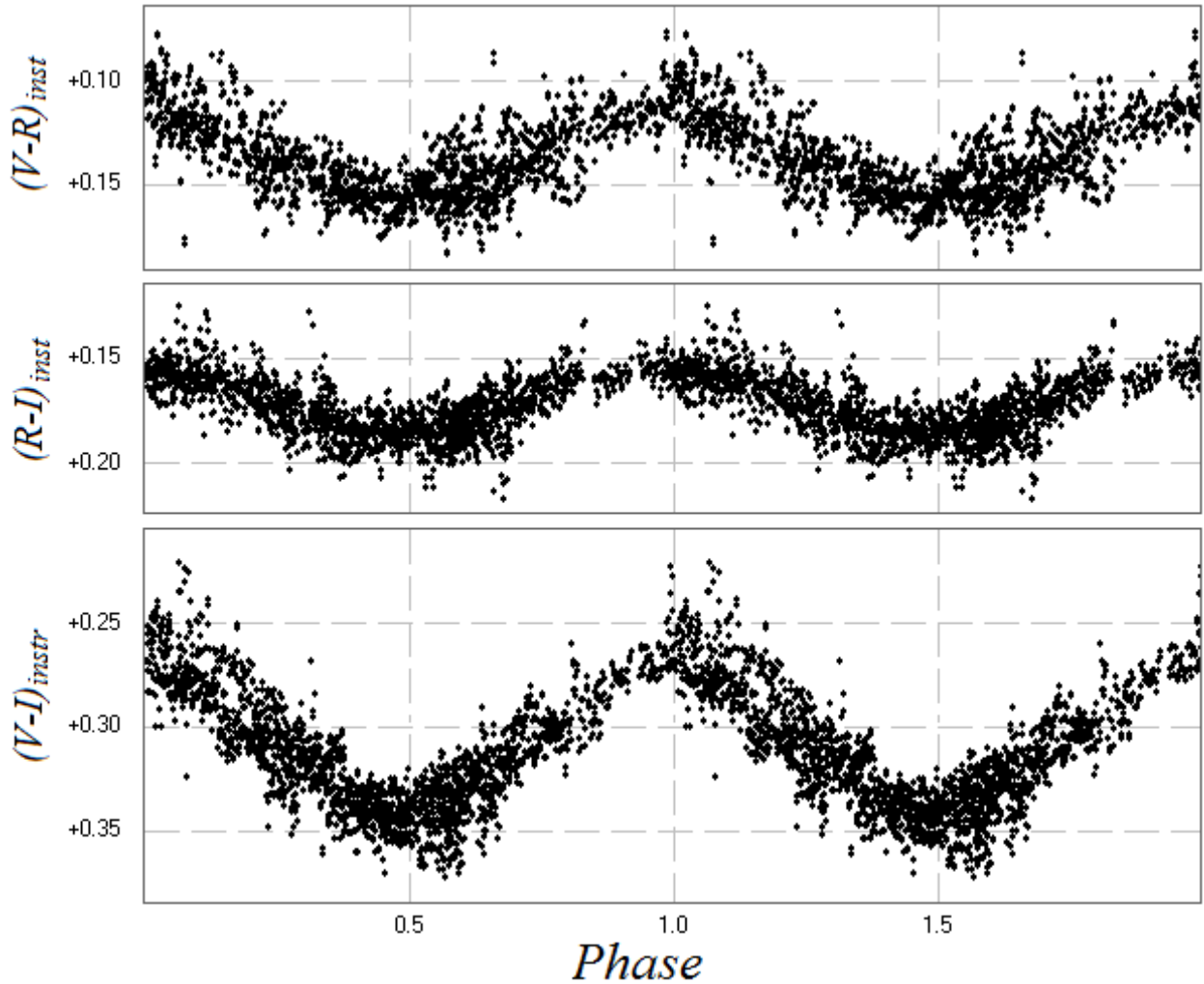

Fig. 6. The phase curves of *V-R*, *R-I* and *V-I* instrumental color indices for CzeV134. I.e. the magnitudes and colors are based on differential photometry using C2 as the main comparison star.

Table 2. Maxima timings and *O-C* for CzeV134.

| HJD 2400000.0+ | error, d | # of cycle | O-C | *Rc* mag | Telescope |
| --- | --- | --- | --- | --- | --- |
| 53236.50412 | 0.00560 | 0 | 0 | 11.502 | SL |
| 53245.34365 | 0.00772 | 21 | 0.02386 | 11.467 | SL |
| 53988.33321 | 0.00552 | 1791 | -0.02197 | 11.490 | SL |
| 53990.45876 | 0.00348 | 1796 | 0.00461 | 11.474 | SL |
| 54335.49031 | 0.00562 | 2618 | -0.03450 | 11.401 | JP |
| 54780.48008 | 0.00734 | 3678 | -0.02638 | 11.470 | JP |
| 55052.51440 | 0.00582 | 4326 | -0.01856 | 11.522 | KS |
| 55058.40716 | 0.00448 | 4340 | -0.00292 | 11.428 | JP |
| 55060.48899 | 0.00357 | 4345 | -0.02006 | 11.428 | JP |

Like in the case of CzeV134, the light curve of this star is variable. In Fig. 7, where the *VRcIc* phase curves of yr. 2004 (SL telescope) are shown, the O'Connell effect is clearly visible. And it's also visible on Fig. 8, where for the phase curve plotting two *V* and *Rc* sets of yr. 2006. observations were used instead of two first series of yr. 2004. However, the asymmetry changed. While on Fig. 7 the second maximum is brighter than the first one and the magnitudes differences in maxima are $(Max_I - Max_{II})_{Ic} = 0.019(2)$ mag, $(Max_I - Max_{II})_{Rc} = 0.022(3)$ mag, $(Max_I - Max_{II})_V = 0.022(3)$ mag, in Fig. 8 *V*-band and *Rc*-band phase curves are nearly symmetric, the first maximum is only slightly brighter than the second one in *Rc*-band: $(Max_I - Max_{II})_{Rc} = -0.009(2)$ mag, $(Max_I - Max_{II})_V = -0.002(3)$ mag.

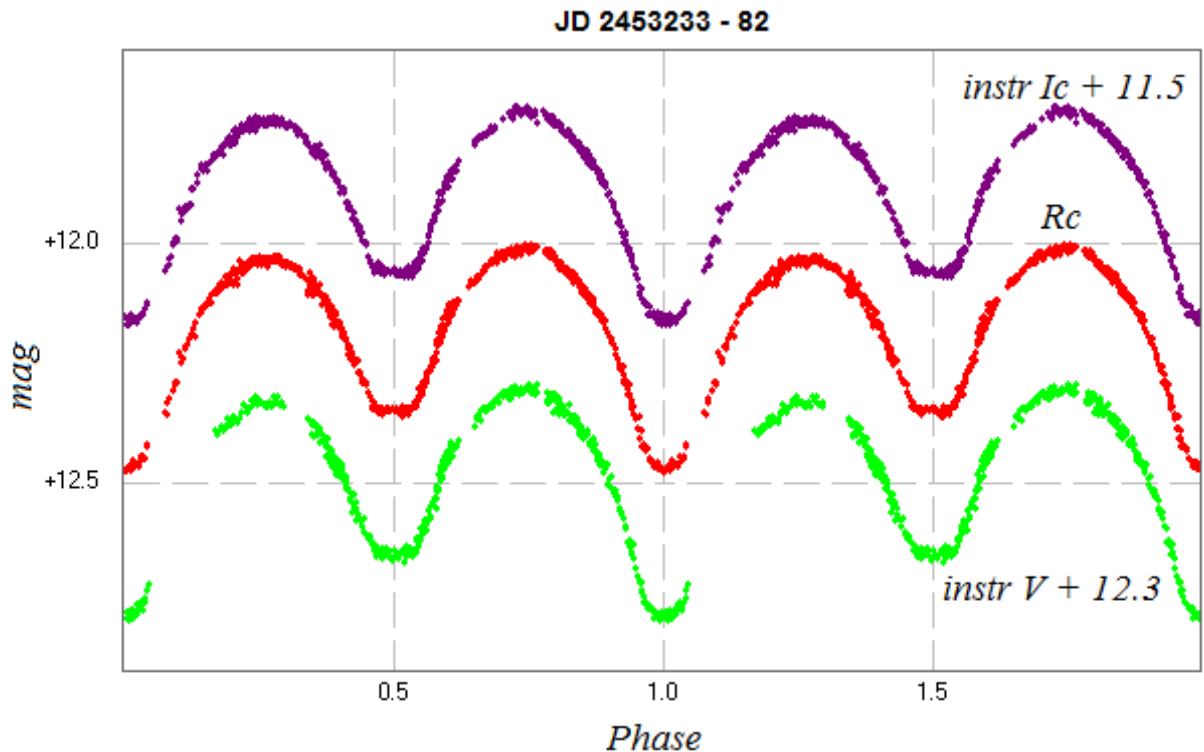

Fig. 7. *V, Rc* and *Ic* phase curves of binary system CzeV135=V1094 Cas for the time interval JD 2453233 – 2453282, plotted using the observations from SL telescope.

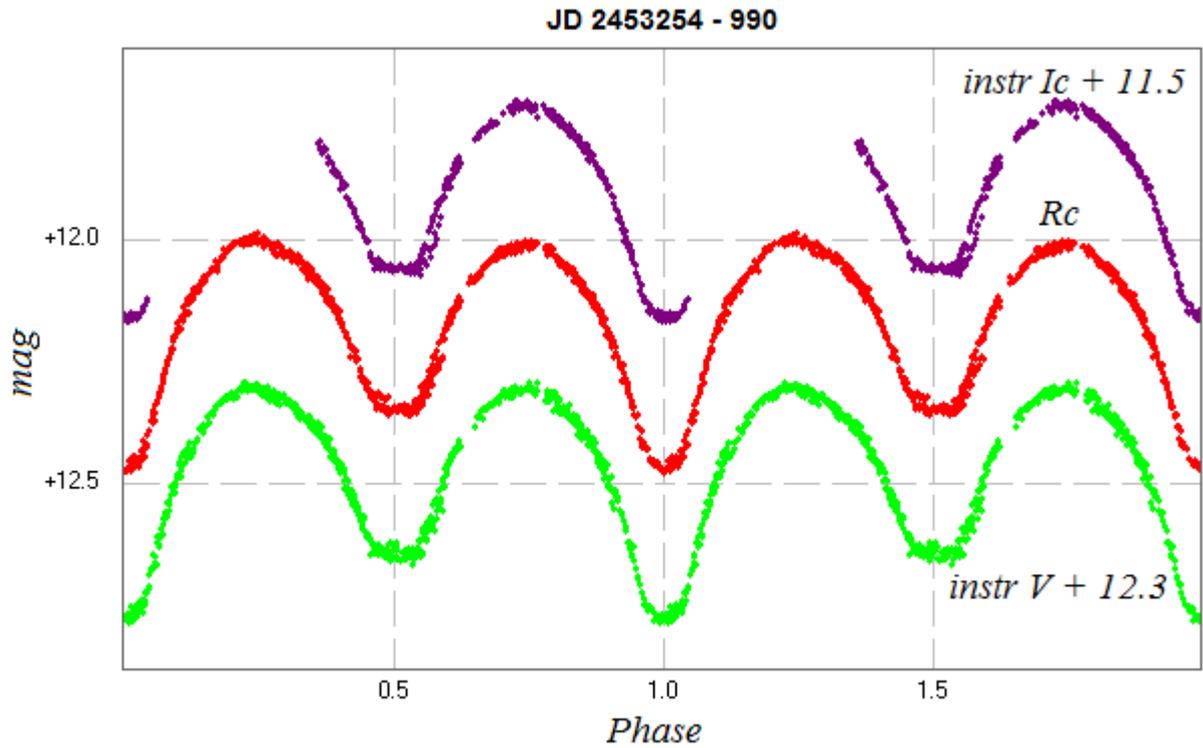

Fig. 8. *V, Rc* and *Ic* phase curves of binary system CzeV135 for the time interval JD 2453254 – 2453990, plotted using the observations from SL telescope.

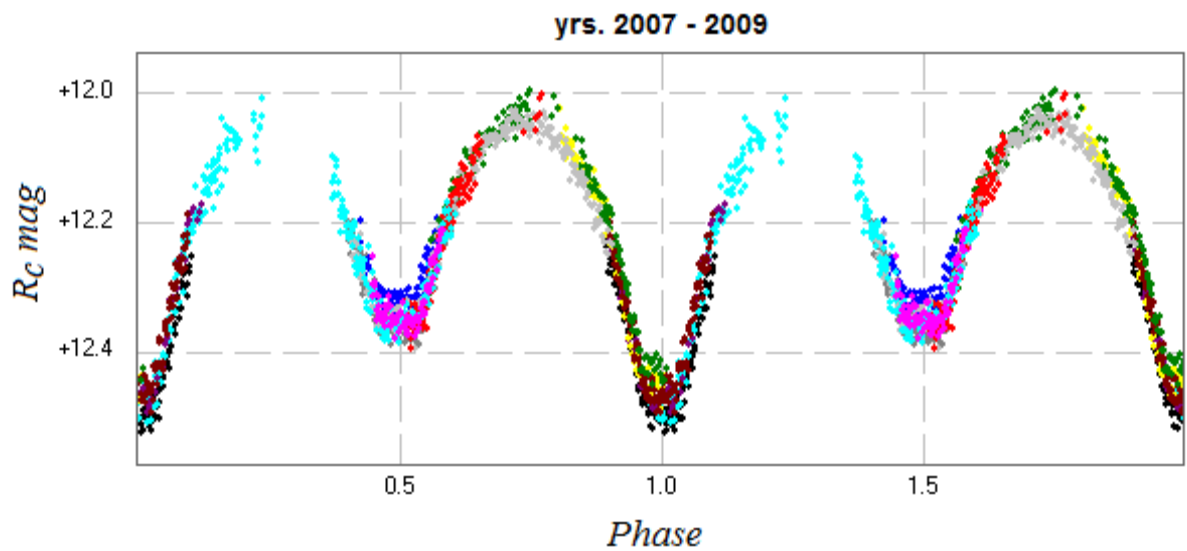

Fig. 9. *Rc*-band phase curve of CzeV135, plotted using the observations from the JP telescope. Different colors correspond to different nights.

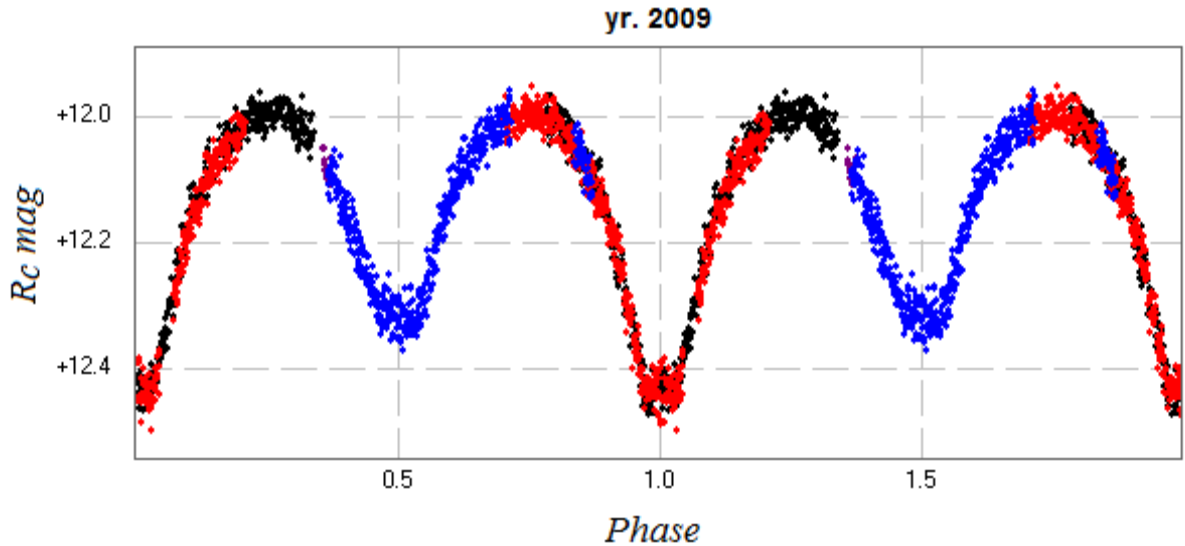

Fig. 10. *Rc*-band phase curve of CzeV135, plotted using the observations from the KS telescope. Different colors correspond to different nights.

The observations from the JP telescope, obtained during yrs. 2007 – 2009 in *Rc*-band, cover nearly whole phase light curve (Fig. 9). The first maximum is covered poorly, thus we cannot say which maximum is brighter. Each night of observations is marked by another color and a slight variability of the light curve could be noticed. On Fig. 10 the phase curve of observations collected with the KS telescope in yr. 2009 is shown. The heights of maxima are equal, but a slight asymmetry of the shape of secondary minimum is visible. The cause of all these changes in the shape of the phase curve could be the spots in the photosphere, which typically undergo changes of their positions, temperatures and radii.

The average brightness in maxima in *Rc*-band is Max = 12.006(4) mag, in minima $\min_I$ = 12.470(4) mag, $\min_{II}$ = 12.350(4) mag. The phase curve corresponds either to EB, or EW type. According to th GCVS phenomenological classification (Samus et al. 2011). In the EB – type systems, the "secondary minimum is observed in all cases, its depth usually being considerably smaller than that of the primary minimum", contrary to the EW-type systems, where the minima "are almost equal or differ insignificantly". However, there was published no numerical value limiting the "significant" and "insignificant" difference in minima. Linnaluoto and Vilhu (1973) studied statistics of 698 β Lyr and W UMa-type eclipsing binaries. The classification parameter $\Delta=\min_I-\min_{II}$ lies in the range 0-0.25 mag for >90% of the EW systems, thus the observed value 0.120(6) mag for CzeV135 is in the middle of this interval. For the EB-type systems, the range is 0-0.85 mag (and even more for few systems), with ~10% systems in a range 0.05-0.15mag. Thus either EB, or EW classifications may be excluded.

Linnaluoto and Vilhu (1973) also introduced another parameter related to mean surface brightnesses of the stars (neglecting limb darkening):

$$F = \frac{10^{0.4(\min_I - \min_{II})} - 1}{10^{0.4(\min_I - \max)} - 1}$$

For CzeV135, *F*=0.219(4), which corresponds to descending and ascending branches of histograms of distributions of EW and EB systems, respectively.

Malkov et al. (2007) proposed a procedure for classification of eclipsing variables. At his diagram (Fig. 1), the system CzeV135 lies close to the theoretical line of equal temperatures, being surrounded by points of DM (detached Main-sequence components) and C (contact) systems. The DM type is less sure, as the orbital period lies outside the range 0.593-30 days for known systems.

Because the flat minimum (full eclipse) is of a smaller depth, the larger star is hotter. This corresponds to a "A-subtype" binary system according to the classification by Binnendijk (1970).

Some additional observations were found in the SuperWASP database (Butters et al, 2010) and the individual minima were determined and used for the $O–C$ analysis. Each primary and secondary minimum was smoothed by the algebraic polynomial of degree $n = 4$, using the least squares method. The list of minima is given in Table 3 in the chronological order. The $O–C$ diagram is plotted in Fig. 11 with corresponding error bars; the linear trend may be suggested. The weighted linear approximation (MCV) is:

$$O–C = -5.6(\pm 3.7) \cdot 10^{-7} \cdot E - 0.00177(\pm 0.00041)$$

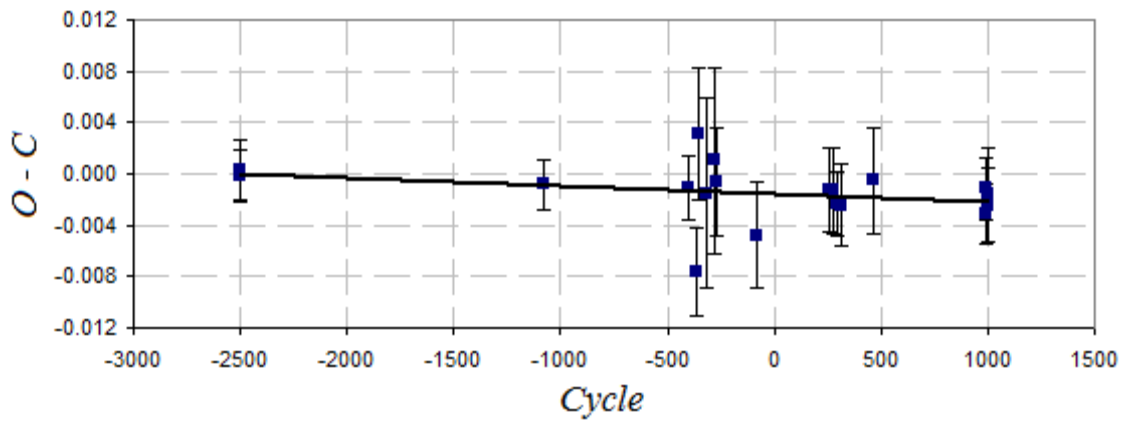

Fig. 11. *O-C* diagram for CzeV135.

The period correction is $4.3\sigma$, thus formally may be suggested to be taken into account, and we determine the new ephemeris:

$$HJD\ T_{\min I} = 2454543.79183 + 0.51429094 \cdot E$$
$$\pm 0.00041 \pm 0.00000037$$

The resulting minima timings are listed in Table 3 together with corresponding errors estimates, $O–C$ differences, and type of minimum (I – primary, II – secondary). Although different methods for time series analysis may lead to slightly other values, we conclude that currently no period change is detected, and an apparent small trend at the O-C diagram may be due to error estimates and possibly changing shape of the phase light curve.

Table 3. Minima timings and *O-C* for CzeV135 (*R*-band)

| HJD, 2400000+ | error | min | #of cycle | O-C | source |
|---|---|---|---|---|---|
| 53256.52230 | 0.00230 | I | -2503 | 0.00032 | SL |
| 53259.35043 | 0.00205 | II | -2497.5 | -0.00015 | SL |
| 53988.35794 | 0.00200 | I | -1080 | -0.00084 | SL |
| 54335.50441 | 0.00248 | I | -405 | -0.00113 | JP |
| 54357.61245 | 0.00344 | I | -362 | -0.00763 | WASP |
| 54363.53756 | 0.00513 | II | -350.5 | 0.00313 | WASP |
| 54381.53313 | 0.00735 | II | -315.5 | -0.0015 | WASP |
| 54398.50727 | 0.00730 | II | -282.5 | 0.00102 | WASP |
| 54405.44852 | 0.00418 | I | -269 | -0.00067 | WASP |
| 54500.33118 | 0.00412 | II | -84.5 | -0.00479 | JP |
| 54676.47958 | 0.00325 | I | 258 | -0.00123 | JP |
| 54683.42245 | 0.00332 | II | 271.5 | -0.00129 | JP |
| 54697.56440 | 0.00249 | I | 299 | -0.00236 | JP |
| 54705.53585 | 0.00325 | II | 314.5 | -0.00243 | JP |
| 54780.62430 | 0.00412 | II | 460.5 | -0.00054 | JP |
| 55052.42478 | 0.00236 | I | 989 | -0.00311 | KS |
| 55054.48392 | 0.00243 | I | 993 | -0.00114 | KS |
| 55059.36841 | 0.00285 | II | 1002.5 | -0.00242 | KS |
| 55060.39783 | 0.00368 | II | 1004.5 | -0.00158 | JP |

**Discussion**

We presented the results of multicolor studying of two recently discovered variable stars: CzeV134 = USNO-A2.0 1425-1870026 and CzeV135 = USNO-A2.0 1425-1825909 = V1094 Cas, and found for them all parameters needed for GCVS.

For the star CzeV134 = USNO-A2.0 1425-1870026 we have refuted the present classification as EW-type, indicating the pulsating nature of variability and discovering changes of the shape of the phase curve. However, there is an ambiguity in the classification of this star. According to the classical types' determination, the stars of RRab subtype have periods from 0.3 to 1.2 days, and amplitudes from 0.5 to 2 mag (V). As the amplitude of CzeV134 is smaller, we can omit it from the RRab subtype. Sometimes the light curve is almost symmetric (the average rising time amounts to about 52% of the period), which matches more likely to RRc subtype. Along with calculated period $P = 0.419782$ d, which is suitable for RRc subtype, we can categorize the CzeV134 like RRc subtype. The photometric parameters (period, amplitude and color index variability) also matches to the β Cep type. However, according to the 2MASS photometric catalog, the color indices of this star are $J - H = 0.364$ mag, $J - K = 0.471$ mag, which corresponds to the G6 spectral class. It matches neither with β Cep, nor with classical description of RRc-type star. The significant changes of the light curve argue for the Blazhko phenomenon in this star. However, from present observations, the period of modulation of the light curve was not found because of too many possible candidate periods. This is a subject for further observational campaign on this star.

Another studied variable, CzeV135 = USNO-A2.0 1425-1825909 = V1094 Cas, is clearly a short-periodic close binary system, from the phenomenological point of view it could be classified as EW-type contact binary. However, taking into account the difference of depth of minima, the EB type formally may not be excluded. This system is interesting because of variability of its phase curve – variable O'Connell effect. We suppose that some kind of chromospheric activity is present. The subtype of this star was recognized as A-subtype. However, the classification is rather unsure, as the shape of the phase curve is under the influence of presence of spots.

In Table 4 all photometric parameters are summarized. For the further more detailed and accurate studying the time-resolved spectral observations are required for both stars.

Table 4. Parameters, needed for the GCVS. The magnitudes are in the *Rc*-band.

| Name | Type | Period, d | Initial epoch, HJD | Max, mag | $Min_I$, mag | $Min_{II}$, mag |
|------|------|-----------|--------------------|----------|--------------|-----------------|
| CzeV134 | RRc | 0.419794(29) | 2453236.50412(56) | 11.457(2) | 11.633(2) | |
| CzeV135 | EW  | 0.51429094(37) | 2454543.79183(41) | 12.006(4) | 12.470(4) | 12.350(4) |

### Acknowledgements

This research has been made using the SIMBAD and VizieR databases operated by the Centre de Données Astronomiques (Strasbourg). This publication makes use of data products from the Two Micron All Sky Survey, which is a joint project of the University of Massachusetts and the Infrared Processing and Analysis Center/California Institute of Technology, funded by the National Aeronautics and Space Administration and the National Science Foundation.

We acknowledge the WASP consortium which comprises of the University of Cambridge, Keele University, University of Leicester, The Open University, The Queen's University Belfast, St. Andrews University and the Isaac Newton Group. Funding for WASP comes from the consortium universities and from the UK's Science and Technology Facilities Council.

The observations in Astronomical Observatory at Kolonica Saddle were done with a support of APVV grant LPP-0024-09.

We also acknowledge overall support and used telescopes with CCD cameras of the Johann Palisa Observatory and Planetarium in Ostrava (http://ostrava.astronomy.cz) and the Astronomical Institute of Slovak Academy of Sciences (http://www.ta3.sk ).